\documentstyle[aps]{revtex}
\begin{document}

\tighten
\twocolumn[\hsize\textwidth\columnwidth\hsize\csname @twocolumnfalse\endcsname

\title{Anomalous Resonance of the Symmetric Single-Impurity Anderson Model in
the Presence of Pairing Fluctuations}

\author{Guang-Ming Zhang$^1$ and Lu Yu$^{2,3}$}

\address{$^1$Center for Advanced Study, Tsinghua University, Beijing 100084,
P. R. of China\\
$^2$International Center for Theoretical Physics, P.O.Box 586, Trieste
34100, Italy\\
$^3$Institute of Theoretical Physics, CAS, Beijing 100080, P.R. of China }

\date{\today}

\maketitle

\begin{abstract}

We consider the symmetric single-impurity Anderson model in the presence of
pairing fluctuations. In the isotropic limit, the degrees of freedom of the
local impurity are separated into hybridizing and non-hybridizing modes. The
self-energy for the hybridizing modes can be obtained exactly, leading to two
subbands centered at $\pm U/2$. For the non-hybridizing modes, the second
order perturbation yields a singular resonance of the marginal Fermi liquid
form. By multiplicative renomalization, the self-energy is derived exactly,
showing the resonance is pinned at the Fermi level, while its strength is
weakened by renormalization.

\end{abstract}

]


The unusual normal state of the high-$T_c$ cuprate superconductors has been
interpreted as a kind of non-Fermi liquid (FL) behavior, but so far there
have been no fully convincing microscopic justifications of such a state in
quasi-two-dimensional systems with short range interactions.
A phenomenological marginal FL
approach was suggested by Varma {\it et al.} \cite{varma} as the ``gentlest''
departure from the conventional FL theory to describe the anomalous
properties. The central hypothesis is that over a wide range of momenta the
charge and spin polarizabilities do not contain any intrinsic low energy
scales, depending only on temperature and external frequency:
\begin{equation}
{\rm Im}\chi _{\rho ,\sigma }(\vec{q},\omega )\sim \left\{
\begin{array}{ll}
-N(0)\text{ }\frac \omega T, & \hspace{0.5cm}\omega \ll T \\
-N(0), & T\ll \omega \ll \omega _c,
\end{array}
\right.
\end{equation}
where $N(0)$ is the density of states at the Fermi level, and $\omega _c$ is
a high-energy cut-off of the order of $0.5$ eV. Apart from cuprates, the
anomalous behavior observed in the normal state of some heavy
fermion superconductors can also be interpreted in terms of such a marginal
FL \cite{maple}. A number of serious attempts have been made to provide a
microscopic justification for this phenomenological theory, beginning with
the three-body resonance \cite{ruckenstein}, followed by a detailed study of
related quantum impurity models\cite{kotliar,perakis,sire,gm,ks}, and the
most recent interpretation is based on local fluctuations of the order
parameter near the quantum critical point using a two-band model for cuprates%
\cite{varma1}. As far as we understand, the issue is still open.

On the other hand, a large body of experimental data in NMR, $\mu $SR,
specific heat, transport, tunneling, neutron scattering, and photoemission
measurements in most underdoped cuprate superconductors can be interpreted
as due to the presence of a pseudo-gap (depleted density of states at the
Fermi level) above $T_c$\cite{randeria,martin,puchkov}. There have been
various proposals for the origin of such a pseudo-gap, and one of them is
due to precursor pairing fluctuations\cite{randeria}. However,
 a possible connection of such pairing fluctuations
with marginal FL behavior has not yet been explored so far.

In this Letter, we, following the experience in the heavy fermion studies,
consider the hybridization of a strongly localized electron with an
uncorrelated
conduction electron band, in terms of an impurity model\cite{anderson}.
Encouraged by the recovering of non-FL behavior in the single-impurity
two-channel Kondo model \cite{ek,coleman}, we explore a generalized
symmetric Anderson model with particle-particle mixing on top of the
standard particle-hole mixing. This model can be justified in the presence
of pairing fluctuations between the local impurity and conduction electrons.
When the two types of hybridizations have the same strengths, the local
impurity degrees of freedom are separated into hybridizing and
non-hybridizing modes, and the anomalous polarizability (1) emerges
naturally as due to the impurity contribution in the non-interacting limit.
The self-energy for the hybridizing modes can be obtained exactly, and the
corresponding density of states forms two subbands centered at $\pm U/2$,
with $U$ as the Hubbard repulsion. Meanwhile, the self-energy for the
non-hybridizing modes is calculated in the second order perturbation, giving
rise to a singular resonance at the Fermi level, of the marginal FL type.
Using the multiplicative renormalization method, the exact self-energy can
be derived self-consistently in the weak coupling limit and extended then to
strong coupling. We find that the non-FL resonance is pinned at the
Fermi level and it is described by the X-ray edge type of singularity.

The symmetric single impurity Anderson model in the presence of pairing
fluctuations is given by:
\begin{eqnarray}
H &=&\sum_{\vec{k},\sigma }\epsilon _{\vec{k}}C_{\vec{k},\sigma }^{\dag }C_{%
\vec{k},\sigma }+\frac V2\sum_{\vec{k},\sigma }(C_{\vec{k},\sigma }^{\dag
}d_\sigma +d_\sigma ^{\dag }C_{\vec{k},\sigma })  \nonumber \\
&&+U(d_{\uparrow }^{\dag }d_{\uparrow }-\frac 12)(d_{\downarrow }^{\dag
}d_{\downarrow }-\frac 12)  \nonumber \\
&&+\frac{V_a}2\sum_{\vec{k},\sigma }(C_{\vec{k},\sigma }^{\dag }d_\sigma
^{\dag }+d_\sigma C_{\vec{k},\sigma }),
\end{eqnarray}
where the symmetric condition $\epsilon _d=-U/2$ has been assumed, and
the chemical potential of conduction electrons is set
to zero. It is worthwhile to note that the $V_a$ term is  similar
to the transverse interaction term for the second channel in the effective
two-channel Kondo Hamiltonian obtained in the abelian bosonization approach
\cite
{ek}. We will focus on the case of $V_a=V$, corresponding to
the isotropic case of the model. As justification of this model we would
mention that Varma has emphasized the importance of the potential scattering
at the impurity $V_0\sum_{k,k^{\prime },\sigma }c_{k,\sigma }^{\dagger
}c_{k^{\prime },\sigma }d_\sigma ^{\dagger }d_\sigma $\cite{varma2}, and the
renormalization group analysis showed that the unitarity limit for this
potential scattering is being approached, while the opposite-spin scattering
is renormalized to zero\cite{gm}. A mean field decoupling of this quartic
term allowing ``anomalous average'' $\langle d_\sigma c_\sigma \rangle $ due
to pairing fluctuations would lead to the particle-particle mixing term in
Eq. (2). Such a pairing is of triplet character, and whether it is related
to the ``odd frequency'' pairing discussed earlier\cite{balatsky} is not
considered here. Nor we are going to discuss the self-consistency issue of
the pairing per se. We will rather focus on the physical consequences of the
above well-defined model.

In order to reveal the characteristics of this generalized model, it is very
useful to introduce the Majorana fermion representation \cite{coleman}: $
d_{\uparrow }=(d_1-id_2)/\sqrt{2}$, \thinspace $d_{\downarrow }=
(-d_3-id_0)/\sqrt{2}$. The model Hamiltonian can then be rewritten
as $H=H_0+H_I$, where
\begin{eqnarray}
&&H_0=\sum_{\vec{k},\sigma }\epsilon _{\vec{k}}C_{\vec{k},\sigma }^{\dag }C_{%
\vec{k},\sigma }  \nonumber \\
&&\qquad \;+\frac V{\sqrt{2}}\sum_{\vec{k}}\left[ (C_{\vec{k},\uparrow
}^{\dag }-C_{\vec{k},\uparrow })d_1-(C_{\vec{k},\downarrow }^{\dag }-C_{\vec{%
k},\downarrow })d_3\right],  \nonumber \\
&&H_I=Ud_1d_2d_3d_0.
\end{eqnarray}
Now it is clear that the pairing fluctuation terms have changed the
structure of the model: one half degrees of freedom of the local impurity ($%
d_1,d_3$) hybridize with the conduction electrons, while another half
degrees of freedom ($d_0,d_2$) decouple. The former will be referred to as
hybridizing modes, whereas the latter as non-hybridizing modes. These two
types of modes are correlated through the Hubbard
interaction. We emphasize that this change is non-perturbative in nature and
makes the non-interacting limit nontrivial.

In the $U=0$ limit, $H_0$ is solved exactly. The Green's functions for the
hybridizing and non-hybridizing modes are calculated as $G_h^{(0)}(\omega
_n)=\frac 1{i\omega _n+i\Delta \text{sgn(}\omega _n\text{)}}$ and \thinspace
$G_{nh}^{(0)}(\omega _n)=\frac 1{i\omega _n},$ where $\omega _n=(2n+1)\pi
/\beta $ with $\beta =1/T$, $\Delta =\pi \rho (0)V^2$ is the hybridization
width, $\rho (0)$ is the density of states of conduction electrons at the
Fermi level, and $\Delta $ will be the high energy cutoff in the weak
coupling theory. The Green's function for the non-hybridizing modes
describes a fermionic zero mode and $G_{nh}^{(0)}(\tau )=-{\rm sgn}(\tau )/2$%
. The total impurity density of states is thus obtained as $A_d^{(0)}(\omega
)=\frac 1\pi \frac \Delta {\omega ^2+\Delta ^2}+\delta (\omega )$. Apart
  from a Lorentz distribution, a delta function peak due to the
non-hybridizing modes appears at the Fermi level. As a result, the
impurity charge and spin density fluctuations will be affected. Since the
charge and spin density operators are defined as
\begin{eqnarray}
n_d &=&\frac 12(d_{\uparrow }^{\dag }d_{\uparrow }+d_{\downarrow }^{\dag
}d_{\downarrow }-1)=-\frac i2(d_1d_2+d_0d_3),  \nonumber \\
S_d^z &=&\frac 12(d_{\uparrow }^{\dag }d_{\uparrow }-d_{\downarrow }^{\dag
}d_{\downarrow })=-\frac i2(d_1d_2-d_0d_3),
\end{eqnarray}
the charge and spin dynamical susceptibilities are both given by
\begin{equation}
\chi _{\rho ,\sigma }^{(0)}(\Omega _n)=\frac 1{2\beta }\sum_{\omega
_n}G_h^{(0)}(\omega _n)G_{nh}^{(0)}(\Omega _n-\omega _n),
\end{equation}
where $\Omega _n=2n\pi /\beta $. After summation over frequency, the
retarded imaginary part is obtained:
\begin{equation}
{\rm Im}\chi _{\rho ,\sigma }^{(0)}(\Omega ,T)=-\frac \pi 4N(\Omega )\tanh
\left( \frac \Omega {2T}\right) ,
\end{equation}
where $N(\Omega )=\frac 1\pi \frac \Delta {\Omega ^2+\Delta ^2}$ is the
density of states of the hybridizing modes. This is exactly what has been
assumed in the marginal FL phenomenology \cite{varma}, of  the same form as
Eq.(1)! It becomes transparent that the anomalous impurity charge and spin
polarizability is due to the delta resonance at the Fermi level, which is
induced by the pairing fluctuations under the condition $V_a=V$. An
analogous situation previously appeared in the studies of the two-channel
Kondo model in the isotropic case \cite{ek,coleman,zh}.

Now consider the effects of interaction on the impurity density of states.
The perturbation theory in terms of Majorana fermions has been fully
developed in the earlier publications \cite{zh}. In the second order
perturbation, the self-energy of the hybridizing modes is given by: $\Sigma
_h(\tau -\tau ^{\prime })=\frac{U^2}4G_h^{(0)}(\tau -\tau ^{\prime }).$ It
is worthwhile to note that this self-energy is {\it exact}, which
can also be obtained by means of retarded double-time Green's functions with
the help of $[d_2d_0,H]=0$. So the exact Green's function is
\begin{equation}
G_h^{-1}(\omega _n)=\left[ G_h^{(0)}(\omega _n)\right] ^{-1}-\frac{U^2}4%
G_h^{(0)}(\omega _n),
\end{equation}
and the density of states Im$G_h(\omega )$ is thus
\begin{equation}
\frac 1{2\pi }\left[ \frac \Delta {(\omega -U/2)^2+\Delta ^2}+\frac \Delta
{(\omega +U/2)^2+\Delta ^2}\right] .
\end{equation}
This spectral density is similar to that of the conventional symmetric
single impurity Anderson model in the Hartree-Fock approximation \cite
{anderson}.

Meanwhile, the self-energy of the non-hybridizing modes in the second order
perturbation is expressed by:
\begin{eqnarray}
&&\Sigma _{nh}^{(2)}(\omega _n)  \nonumber \\
&&\hspace{0.3cm}=\frac{U^2}{\beta ^2}\sum_{\omega _1,\omega
_2}G_h^{(0)}(\omega _1)G_h^{(0)}(\omega _2)G_{nh}^{(0)}(\omega _n-\omega
_1-\omega _2).
\end{eqnarray}
Using the spectral representation, the hybridizing Green's functions are
expressed in terms of their density of states $N(\epsilon )$, and after the
frequency summation we find
\begin{eqnarray}
\Sigma _{nh}^{(2)}(\omega _n) &=&\frac{U^2}4\int_{-\infty }^\infty d\epsilon
_1d\epsilon _2\frac{N(\epsilon _1)N(\epsilon _2)}{i\omega _n-\epsilon
_1-\epsilon _2}\tanh \left( \frac{\epsilon _1}{2T}\right)   \nonumber \\
&&\qquad \quad \text{\quad }\times \left[ \tanh \left( \frac{\epsilon _2}{2T}%
\right) +\coth \left( \frac{\epsilon _1}{2T}\right) \right] .
\end{eqnarray}
Carrying out the analytical continuation $i\omega _n\rightarrow \omega $ and
approximating $N(\epsilon )$ by its value at the Fermi level, the imaginary
part of the retarded self-energy is derived as
\begin{equation}
{\rm Im}\Sigma _{nh}^{(2)}(\omega ,T)=-\frac \pi 2\left( \frac U{\pi \Delta }%
\right) ^2\omega \coth \left( \frac \omega {2T}\right) ,
\end{equation}
and the corresponding real part is also obtained as ${\rm Re}\Sigma
_{nh}^{(2)}(\omega ,T)\approx \left( \frac U{\pi \Delta }\right) ^2\omega
\ln \left( \frac{{\rm max}(\omega ,T)}\Delta \right) .$ This self-energy
gives rise to a quasiparticle weight depending logarithmically on frequency
or temperature, namely,
\begin{equation}
z_{nh}^{(2)}\approx 1+\left( \frac U{\pi \Delta }\right) ^2\ln \left( \frac{%
{\rm max}(\omega ,T)}\Delta \right) .
\end{equation}
In the corresponding Green's function, the pole $\omega =0$ is preserved and
the resonance is characterized by a marginal FL origin. To some extent, the
non-hybridizing modes may represent the low energy excitations of the model.

Since there is a logarithmic correction to the self-energy for the
non-hybridizing modes, we have to examine the higher order expansions
carefully. Moreover, the logarithmic singularity in the reducible interaction
vertex diagrams can be proved to cancel each other, so one
has to consider the  $irreducible$ interaction vertex $\Gamma
_{1,2,3,0}(0,\omega _n;0,\omega _n)\equiv \Gamma (\omega _n)$, which can not
be severed into separate diagrams by cutting a pair of lines, one of which
is the propagator of $G_{nh}$ and the other is $G_h.$ The lowest
order correction to $\Gamma (\omega _n)$ is given by
\[
\frac{U^3}{\beta ^2}\sum_{\omega _1,\omega _2}G_h^{(0)}(\omega
_1)G_h^{(0)}(\omega _2)\left[ G_{nh}^{(0)}(\omega _n-\omega _1-\omega
_2)\right] ^2.
\]
It is interesting to note that this lowest order correction can be related
to $\Sigma _{nh}^{(2)}(\omega _n)$ by the relation $\Gamma ^{(3)}(\omega
_n)=-U\frac{\partial \Sigma _{nh}^{(2)}(\omega _n)}{\partial (i\omega _n)}.$
Therefore, the correction to the  irreducible vertex is also
logarithmically divergent:
\begin{equation}
\Gamma ^{(3)}(\omega ,T)\approx -U\left( \frac U{\pi \Delta }\right) ^2\ln
\left( \frac{\max (\omega ,T)}\Delta \right) .
\end{equation}
In fact, from a general consideration of the model, a Ward
identity can be established \cite{zhang}, relating the  irreducible
vertex to the non-hybridizing self-energy:
\begin{equation}
\Gamma (\omega _n)=U\left[ 1-\frac{\partial \Sigma _{nh}(\omega _n)}{%
\partial (i\omega _n)}\right] ,
\end{equation}
which shows that to a large extent the singularity in $\Gamma (\omega _n)$
and $\Sigma _{nh}(\omega _n)$ cancel each other. This implies that the
renormalization of the self-energy in the higher order expansions must be
treated on an equal footing with that of the irreducible vertex.

In the
treatment of logarithmic problems, the multiplicative renormalization is an
effective method \cite{solyom}.
In such a treatment, the temperature is fixed at zero for simplicity and
only the frequency variables are retained. Since the interaction is cut off
at $\Delta $ in our weak coupling theory, the Green's functions and
vertices will depend only on the ratio $\frac \omega \Delta $, and the
dimensionless coupling constant $\tilde{U}\equiv $ $\frac U{\pi \Delta }$
must be defined as the proper invariant coupling constant. When we express
the  irreducible vertex by $\frac 1{\pi \Delta }\Gamma \equiv \tilde{U%
}\tilde{\Gamma}$, it follows from the structure of the Dyson equation that
the renormalization equations will have the forms
\begin{eqnarray}
&&G_h(\omega ,\Delta ^{\prime },\tilde{U}^{\prime })=z_1G_h(\omega ,\Delta ,%
\tilde{U}),  \nonumber \\
&&G_{nh}(\omega /\Delta ^{\prime },\tilde{U}^{\prime })=z_2G_{nh}(\omega
/\Delta ,\tilde{U}),  \nonumber \\
&&\tilde{\Gamma}(\omega /\Delta ^{\prime },\tilde{U}^{\prime })=z_3^{-1}%
\tilde{\Gamma}(\omega /\Delta ,\tilde{U}),  \nonumber \\
&&\tilde{U}^{\prime }=z_1^{-1}z_2^{-1}z_3\tilde{U}.
\end{eqnarray}
Since its second order perturbational result is exact, the Green's function
for the hybridizing modes should remain unrenormalized and therefore $z_1=1$%
. Using the low order perturbation results for $z_{nh}$ and $\tilde{\Gamma}$%
, we deduce
\begin{equation}
z_2^{-1}=1+\tilde{U}^2\ln \left( \frac{\Delta ^{\prime }}\Delta \right) ,%
\text{ }z_3=1-\tilde{U}^2\ln \left( \frac{\Delta ^{\prime }}\Delta \right) ,
\end{equation}
and $\tilde{U}^{\prime }=\tilde{U}$. The invariant coupling constant is not
renormalized, and this conclusion is actually valid beyond the
multiplicative renormalization approximation, because it is ensured by the
Ward identity!

Meanwhile, the quasiparticle weight $z_{nh}(\omega )$ is also a proper
quantity satisfying the criterion of multiplicative renormalization, and has
good transformation properties. When the second order perturbation result of
$z_{nh}$ is taken into account, the Lie equation up to the second order can
be derived as follows:
\begin{equation}
\frac{\partial \ln z_{nh}}{\partial \ln \omega }=\tilde{U}^2\frac 1\omega .
\end{equation}
By integrating over $\omega $ and determining the constant of integration
  from fitting to the second order perturbation expression, we get $%
z_{nh}(\omega )=\left( \frac \omega \Delta \right) ^{\tilde{U}^2}.$ The
Green's function for the non-hybridizing modes is thus obtained
\begin{equation}
G_{nh}(\omega )=\frac 1\omega \text{ }\left( \frac \omega \Delta \right) ^{%
\tilde{U}^2}.
\end{equation}
To calculate the spectral density Im$G_{nh}$ we note that the branch cut of $%
G_{nh}$ extends only on one side of the branching point. Thus, Im$G_{nh}$ is
a step function of the energy. The phase of the factor $\omega ^\alpha $ is
then completely determined: on the branch cut side it involves a factor $%
(-1)^\alpha .$ In the weak coupling limit $\tilde{U}<<1,$ we thus obtain
\begin{equation}
\text{Im}G_{nh}(\omega )=\pi \tilde{U}^2\frac 1\omega \text{ }\left( \frac %
\omega \Delta \right) ^{\tilde{U}^2}.
\end{equation}
The singularity of the resonance at $\omega =0$ is weakened by
renormalization, and follows a power law with an exponent proportional to
the squared interaction strength. The non-FL quasiparticle weight vanishes
at the Fermi level, thus there are strong similarities
between the present singular resonance and the X-ray emission spectra \cite
{nozieres}, and the marginal FL resonance is actually described by a power law
behavior in the low energy limit. Paralleling the asymptotically exact
calculation for the X-ray edge problem\cite{nd}, we believe that it is
possible
to develop an exact treatment of this resonance for arbitrary coupling
parameter
$\tilde{U}$. This way, the above weak coupling results may be
extended to strong coupling limit after replacing $\tilde{U}$ by $
\delta/\pi,$ where $\delta =\arctan ({{\tilde{U}}/\Delta })$ is an
s-wave phase shift at the Fermi level.

In addition, when the large $U$ limit is considered, a Schrieffer-Wolf
transformation can be applied to the model Hamiltonian, generating an s-d type
of model:
\begin{equation}
 H'=\sum_{\vec{k},\sigma}\epsilon_{\vec{k}}C^{\dag}_{\vec{k},\sigma}
    C_{\vec{k},\sigma}+\frac{4V^2}{U}[\sigma^y(0)+\tau^y(0)]S^y_d,
\end{equation}
where $\sigma^y(0)=-i(C^{\dag}_{0,\uparrow}C_{0,\downarrow}-
       C^{\dag}_{0,\downarrow}C_{0,\uparrow})/2$, and
      $\tau^y(0)=-i(C^{\dag}_{0,\uparrow}C^{\dag}_{0,\downarrow}-
       C_{0,\downarrow}C_{0,\uparrow})/2$, are the y-components of the spin
and isospin densities of the conduction electrons at the impurity site.
$S_d^y=-id_3d_1$ is the corresponding impurity spin operator. This
resulting model is reminiscent of the so-called compactified two-channel Kondo
model \cite{coleman}. However, here is only one component in the exchange
interactions, which become a genuinely $marginal$ operator, being
compatible with the
renormalization analysis of the Hubbard interaction.

The total density of states of the local impurity is a sum of the
hybridizing and non-hybridizing modes: Im$G_h+$ Im$G_{nh}$. Within the
present exact theory, a very interesting picture emerges in a nature way:
The hybridizing modes form two subbands with Lorentz form centered at $\pm
U/2,$ in some sense similar to the lower and upper Hubbard bands, while the
non-hybridizing modes form a singular non-FL resonance at the Fermi level.
The low-energy excitations and high-energy excitations are thus separated,
and only the former is strongly renormalized. More
interestingly, a very similar picture with three peaks in the density of
states has been suggested by Kotliar, Georges, and coworkers\cite{xyz} in
their studies of the symmetric $d=\infty $ one-band Hubbard model close to
the metal-insulator transition, where the central peak is a FL quasiparticle
resonance. It is very likely that the present model including
the pairing fluctuations between the local impurity and conduction electrons
could shed some light on the nature of some  strongly correlated electron
systems.

In conclusion, we have considered the effect of pairing fluctuations on the
symmetric single impurity Anderson model. The main features of this model
are very similar to those anticipated for the ``three body'' resonance
model, considered earlier\cite{ruckenstein}. However, the difficulty due to
generating a new low-energy scale, inherent in the previous model, does not
appear here. This model study suggests a possible microscopic origin of the
marginal FL behavior. After completing the present calculation, we saw a new
preprint\cite{ho} where an attempt was made to make connection between the
marginal FL behavior and a lattice of three-body bound states. However, the
issue considered there is different from ours.

{\bf Acknowledgments}

One of the authors (G.-M Zhang) would like to express his gratitude to
Professor A.C. Hewson for his hospitality at Imperial College of London,
where this work was initiated. L. Yu would like to thank A. Nersesyan
for very helpful discussions.

\end{document}